\documentclass[letterpaper,11pt]{article}
\pdfoutput=1 

\usepackage{jheppub} 

\title{Bi-Local Holography in the SYK Model: Perturbations}

\author{Antal Jevicki,}
\author{Kenta Suzuki}

\affiliation{Department of Physics, Brown University,\\
182 Hope Street, Providence, RI 02912, U.S.A.}

\emailAdd{antal\_jevicki@brown.edu}\emailAdd{kenta\_suzuki@brown.edu}

\preprint{{\tt BROWN-HET-1676}}

\abstract{
We continue the study of the Sachdev-Ye-Kitaev model in the Large $N$ limit.
Following our formulation in terms of bi-local collective fields with dynamical reparametrization symmetry, we perform perturbative calculations around the conformal IR point.
}

\begin{document}
\maketitle
\flushbottom

\section{Introduction}
\label{sec:intro}
In this paper we continue the development of the Large $N$ formulation of the Sachdev-Ye-Kitaev (SYK) model begun in our earlier work.
The SYK model \cite{Kitaev:2015,Kitaev:2014,Sachdev:2015efa,Fu:2016yrv}
and the earlier Sachdev-Ye (SY) model \cite{Sachdev:1992fk,Georges:1999,Sachdev:2010um,Sachdev:2010uj}
represent valuable laboratories for understanding of holography and quantum features of black holes.
They represent fermionic systems with quenched disorder with nontrivial properties \cite{Erdmenger:2015xpq,Anninos:2016szt, Danshita:2016xbo,Garcia-Alvarez:2016wem} and gravity duals.
In addition to models based on random matrices,
they represent some of the simplest models of holography (see also \cite{Hartnoll:2016mdv}).
The framework for accessing the IR critical point and the corresponding AdS$_2$ dual can be provided by the large $N$ expansion at strong coupling.
In this limit, Kitaev \cite{Kitaev:2015} has demonstrated the chaotic behavior of the system 
in terms of the Lyapunov exponent and has exhibited elements of the dual black hole.

Recently, in-depth studies \cite{Polchinski:2016xgd,Jevicki:2016bwu,Maldacena:2016hyu} have given large $N$ correlations and spectrum of two-particle states of the model.
In these (and earlier works \cite{Kitaev:2015,Sachdev:2015efa}), a notable feature is the emergence of reparametrization symmetry showing characteristic features of the dual AdS Gravity.

The present work continues the development of systematic Large $N$ representation of the model given in \cite{Jevicki:2016bwu} (which we will refer to as I), through a nonlinear bi-local collective field theory.
This representation systematically incorporates arbitrary $n$-point bi-local correlators through a set of $1/N$ vertices and propagator(s) and as such
gives the bridge to a dual description. It naturally provides a holographic interpretation along the lines proposed more generally in \cite{Das:2003vw,Koch:2014mxa}, where the relative coordinate is seen to represent the radial AdS$_2$ coordinate $z$. 
The Large $N$ SYK model represents a highly nontrivial nonlinear system. At the IR critical point (which is analytically accessible)
there appears a zero mode problem which at the outset prevents a perturbative expansion. In (I), this is treated through introduction of collective `time' coordinate as a dynamical variable as in quantization of extended systems \cite{Gervais:1975pa}.
Its Faddeev-Popov quantization was seen to systematically project out the zero modes, providing for a well defined propagator and expansion around the IR point.
What one has is a fully nonlinear interacting system of bi-local matter with a discrete gravitational degree of freedom governed by a Schwarzian action.
In \cite{Maldacena:2016hyu} the zero modes were enhanced away from the IR defining a near critical theory, and correspondence.
We will be able to demonstrate that the nonlinear treatment that we employ leads to very same effects (`big' contributions) at the linearized quadratic level, it is expected hoverer to be exact at all orders. 

In the present work, we present perturbative calculations (around the IR point) using this collective formulation.
These calculations are compared with and are seen to be in agreement with numerical evaluations of \cite{Maldacena:2016hyu}. 
The content of this paper is as follows: In the rest of Section \ref{sec:intro}, we give a short summary of our formulation with the treatment of symmetry modes.
In Section \ref{sec:shift of the classical solution}, we perform a perturbative evaluation of the Large $N$ classical background, to all orders in the inverse of the strong coupling defining the IR.
In Section \ref{sec:two-point function}, we discuss the two-point function in the leading and sub-leading order.
In Section \ref{sec:finite temperature}, we deal with the finite temperature case and give the free energy to several orders.
Comments are given in Section \ref{sec:conclusion}.

\subsection{The method}
\label{sec:the method}
In this subsection, we will give a brief review of our formalism \cite{Jevicki:2016bwu}.
The Sachdev-Ye-Kitaev model \cite{Kitaev:2015} is a quantum mechanical many body system with all-to-all interactions on fermionic $N$ sites ($N \gg 1$), represented by the Hamiltonian
	\begin{equation}
		H \, = \, \frac{1}{4!} \sum_{i,j,k,l=1}^N J_{ijkl} \, \chi_i \, \chi_j \, \chi_k \, \chi_l \, ,
	\end{equation}
where $\chi_i$ are Majorana fermions, which satisfy $\{ \chi_i, \chi_j \} = \delta_{ij}$.
The coupling constant $J_{ijkl}$ are random with a Gaussian distribution.
The original model is given by this four-point interaction; however, with a simple generalization to analogous $q$-point interacting model \cite{Kitaev:2015,Maldacena:2016hyu}.
In this paper, we follow the more general $q$ model, unless otherwise specified.
Nevertheless, our main interest represents the original $q=4$ model.
After the disorder averaging for the random coupling $J_{ijkl}$, there is only one effective coupling $J$ and the effective action is written as
	\begin{equation}
		S_q \, = \, - \, \frac{1}{2} \int dt \sum_{i=1}^N \sum_{a=1}^n \chi_i^a \partial_t \chi_i^a
		\, - \, \frac{J^2}{2qN^{q-1}} \int dt_1 dt_2 \sum_{a, b=1}^n \left( \sum_{i=1}^N \chi_i^a(t_1) \chi_i^b(t_2) \right)^q \, ,
	\end{equation}
where $a, b$ are the replica indexes. Throughout this paper, we only use Euclidean time.
We do not expect a spin glass state in this model \cite{Sachdev:2015efa} and we can restrict to replica diagonal subspace \cite{Jevicki:2016bwu}.
Therefore, introducing a (replica diagonal) bi-local collective field:
	\begin{equation}
		\Psi(t_1, t_2) \, \equiv \, \frac{1}{N} \sum_{i=1}^N \chi_i(t_1) \chi_i(t_2) \, ,
	\end{equation}
the model is described by a path-integral
	\begin{equation}
		Z \, = \, \int \prod_{t_1, t_2} \mathcal{D}\Psi(t_1, t_2) \ \mu(\Psi) \, e^{-S_{\rm col}[\Psi]} \, , 
	\label{eq:collective partition function}
	\end{equation}
with an appropriate order $\mathcal{O}(N^0)$ measure $\mu$ and the collective action:
	\begin{equation}
		S_{\rm col}[\Psi] \, = \, \frac{N}{2} \int dt \, \Big[ \partial_t \Psi(t, t')\Big]_{t' = t} \, + \, \frac{N}{2} \, {\rm Tr} \log \Psi \, - \, \frac{J^2N}{2q} \int dt_1 dt_2 \, \Psi^q(t_1, t_2) \, ,
	\label{S_col}
	\end{equation}
where the trace term comes from a Jacobian factor due to the change of path-integral variable, and the trace is taken over the bi-local time.
This action being of order $N$ gives a systematic $G=1/N$ expansion, while the measure $\mu$ found as in \cite{Jevicki:2014mfa} begins to contribute at one loop level (in $1/N$).
Here the first linear term represents a conformal breaking term,
while the other terms respect conformal invariance.\footnote{Such linear breaking term was seen previously in \cite{Read:1995}.}
This naive expression of the breaking term represents a product at the same point, which will be receiving regularization in our perturbation.
In the IR with the strong coupling  $|t|J \gg 1$, the collective action is reduces to the critical action 
	\begin{equation}
		S_{\rm c}[\Psi] \, = \, \frac{N}{2} \, {\rm Tr} \log \Psi \, - \, \frac{J^2N}{2q} \int dt_1 dt_2 \, \Psi^q(t_1, t_2) \, ,
	\label{S_c}
	\end{equation}
which exhibits the emergent conformal reparametrization symmetry $t \to f(t)$ with
	\begin{equation}
		\Psi(t_1, t_2) \, \to \, \Psi_f(t_1, t_2) \, = \, \Big| f'(t_1) f'(t_2) \Big|^{\frac{1}{q}} \, \Psi(f(t_1), f(t_2)) \, .
	\label{reparametrization}
	\end{equation}
The critical solution is given by
	\begin{equation}
		\Psi_{0, f}(t_1, t_2) \, = \, b \left( \frac{\sqrt{|f'(t_1)f'(t_2)|}}{|f(t_1) -f(t_2)|} \right)^{\frac{2}{q}} \, ,
	\label{Psi_{0,f}}
	\end{equation}
where $b$ is a time-independent constant.	
This symmetry is responsible for the appearance of zero modes in the strict IR critical theory.
This problem was addressed in \cite{Jevicki:2016bwu} with analog of the quantization of extended systems with symmetry modes \cite{Gervais:1975pa}.
The above symmetry mode representing time reparametrization can be elevated to a dynamical variable introduced according to \cite{Gervais:1975yg}
through the Faddeev-Popov method which we summarize as follows: we insert into the partition function (\ref{eq:collective partition function}), the functional identity:
	\begin{equation}
		\int \prod_{t} \mathcal{D}f(t) \ \prod_{t}\delta\left( \int u \cdot \Psi_f \right) \left| \frac{\delta \left(\int u \cdot \Psi_f \right)}{\delta f} \right| \, = \, 1 \, ,
	\end{equation}
so that after an inverse change of the integration variable, it results in a combined representation 
	\begin{equation}
		Z \, = \, \int \prod_{t} \mathcal{D}f(t) \prod_{t_1, t_2} \mathcal{D}\Psi(t_1, t_2) \ \mu(f, \Psi) \, \delta\left( \int u \cdot \Psi_f \right) \, e^{-S_{\rm col}[\Psi, f]} \ ,
		\label{eq:gauged collective partition function}
	\end{equation}
with an appropriate Jacobian.
After separating the critical classical solution $\Psi_0$ from the bi-local field: $\Psi=\Psi_0+\overline{\Psi}$, the total action is now given by
	\begin{equation}
		S_{\rm col}[\Psi, f] \, = \, S[f] \, + \, \frac{N}{2} \int \big[ \overline{\Psi}_f \big]_s \, + \, S_{\rm c}[\Psi] \, .
	\label{S_col[Psi, f]}
	\end{equation}
Here [ ]$_s$ represents a regularized expression for the breaking operator, that we will specify in Section \ref{sec:evaluation of psi_1}.
The action of the time collective coordinate is given by
	\begin{equation}
		S[f] \, = \, \frac{N}{2} \int \, \big[ \Psi_{0, f} \big]_s \, .
	\label{S[f] original}
	\end{equation}
We have in ( I ) given the explicit evaluation of the nonlinear action $S[f]$ for the case of $q=2$ 
demonstrating the Schwarzian form \cite{Jevicki:2016bwu} conjectured by Kitaev and constructed at quadratic level by Maldacena and Stanford \cite{Maldacena:2016hyu}.
For general $q$, the naive form of the composite operator in (\ref{S_col}) generates again a Schwarzian action,
which we exhibited through an $\varepsilon$-expansion presented in Appendix \ref{app:epsilon-expansion}.
Taking into account the regularized breaking term we confirm the Schwarzian form (in Appendix \ref{app:s-regularization and schwarzian action})
	\begin{equation}
		S[f] \, = \, - \, \frac{N\alpha}{24\pi J} \int dt \, \left[ \, \frac{f'''(t)}{f'(t)} \, - \, \frac{3}{2} \, \left( \frac{f''(t)}{f'(t)} \right)^2 \, \right] \, ,
	\label{S[f]}
	\end{equation}
with a coefficent 
	\begin{equation}
		\alpha \, = \, - 12 \pi B_1 \gamma \, ,
	\end{equation}
where 
	\begin{equation}
		\gamma(q) \, = \, - \, \frac{\tan(\frac{\pi}{q})}{12\pi b q} \, \left[ \frac{2\pi(q-1)(q-2)}{q\sin(\frac{2\pi}{q})} \, - \, (q^2-6q+6) \right] \, .
	\label{gamma}
	\end{equation}
and $B_1$ representing the coefficient of first order shift of the saddle-point solution which will be summarized in Section \ref{sec:evaluation of psi_1}.
All together our improved result for the prefactor of the Schwarzian action comes out in agreement with the value obtained first by Maldacena and Stanford
through evaluation of zero mode dynamics \cite{Maldacena:2016hyu}.

Summarizing in the above construction we have an interacting picture of the emergent Schwarzian mode $f(t)$,
and a bi-local matter field combined in the nonlinear collective action (\ref{S_col[Psi, f]}).
It is important to emphasize that this action exhibits reparametrization symmetry both at and also away from the IR point.
For this, the delta constraint condition projecting out the state associated with wave function $u(t_1, t_2)$ represents a gauge fixing condition with an corresponding Faddeev-Popov measure.
This formulation then allows systematic perturbative calculations around the IR point.

\subsection{Relation to Zero Mode Dynamics}
\label{sec:relation to zero mode dynamics}
Before we proceed with our perturbative calculations it is worth comparing the above exact treatment of the reparametrization mode (\ref{S[f]}) 
with a linearized determination of the zero mode dynamics, as considered in \cite{Maldacena:2016hyu}. We will be able to see that the latter 
follows from the former.

Expanding the critical action around the critical saddle-point solution $\Psi_0$, we have in I generated \cite{Jevicki:2016bwu},
the quadratic kernel (which defines the propagator) and a sequence of higher vertices.
This expansion is schematically written as
	\begin{equation}
		S_{\rm c}\Big[\Psi_0+\sqrt{2/N} \, \eta\Big] \, = \, N \, S_{\rm c}[\Psi_0] \, + \, \frac{1}{2} \int \eta \cdot \mathcal{K} \cdot \eta
		\, + \, \frac{1}{\sqrt{N}} \int \mathcal{V}_{(3)} \cdot \eta \, \eta \, \eta \, + \, \cdots \, ,
	\end{equation}
where the kernel is
	\begin{align}
		\mathcal{K}(t_1, t_2; t_3, t_4) \, &= \, \frac{\delta^2 S_c[\Psi_0]}{\delta\Psi_0(t_1, t_2) \delta\Psi_0(t_3, t_4)} \nonumber\\
		&= \, \Psi_0^{-1}(t_1, t_3) \Psi_0^{-1}(t_2, t_4) \, + \, (q-1) J^2 \, \delta(t_{13}) \delta(t_{24}) \, \Psi_0^{q-2}(t_1, t_2) \, ,
	\label{kernel}
	\end{align}
with $t_{ij}=t_i-t_j$.
For other detail of the expansion, please refer to \cite{Jevicki:2016bwu}.
Then, the bi-local propagator $\mathcal{D}$ is determined as a solution of the following Green's equation:
	\begin{equation}
		\int dt_3 dt_4 \, \mathcal{K}(t_1, t_2; t_3, t_4) \, \mathcal{D}(t_3, t_4; t_5, t_6) \, = \, \delta(t_{15}) \delta(t_{26}) \, .
	\label{Green's eq}
	\end{equation}
In order to inverse the kernel $\mathcal{K}$ in the Green's equation (\ref{Green's eq}) and determine the bi-local propagator, 
let us first consider an eigenvalue problem of the kernel $\mathcal{K}$:
	\begin{equation}
		\int dt_3 dt_4 \, \mathcal{K}(t_1, t_2; t_3, t_4) \, u_{n, t}(t_3, t_4) \, = \, k_{n,t} \, u_{n,t}(t_1, t_2) \, ,
	\label{eigenvalue problem}
	\end{equation}
where $n$ and $t$ are labels to distinguish the eigenfunctions.
The zero mode, whose eigenvalue is $k_0=0$ is given by
	\begin{equation}
		u_{0,t}(t_1, t_2) \, = \, \frac{\delta \Psi_{0, f}(t_1, t_2)}{\delta f(t)} \bigg|_{f(t)=t} \, .
	\label{zero mode}
	\end{equation}
Now, we consider the zero mode quantum fluctuation around a shifted classical background
	\begin{equation}
		\Psi(t_1, t_2) \, = \, \Psi_{\rm cl}(t_1, t_2) \, + \, \int dt' \, \varepsilon(t') \, u_{0, t'}(t_1, t_2) \, ,
	\end{equation}
with $\Psi_{\rm cl} = \Psi_0 + \Psi_1$ where $\Psi_1$ is a shift of the classical field from the critical point.
Then, the quadratic action of $\varepsilon$ in the first order of the shift is given by expanding $S_{\rm c}[\Psi_{\rm cl}+\varepsilon \cdot u_0]$.
This quadratic action can be written in terms of the shift of the kernel $\delta\mathcal{K}$ as
	\begin{equation}
		S_2[\varepsilon]
		\, = \, - \, \frac{N}{4} \int dt dt' \, \varepsilon(t) \, \varepsilon(t') \int dt_1 dt_2 dt_3 dt_4 \, u_{0, t}(t_1, t_2) \, \delta\mathcal{K}(t_1, t_2; t_3, t_4) \, u_{0, t'}(t_3, t_4) \, ,
	\label{S_2[epsilon]1}
	\end{equation}
where
	\begin{equation}
		\delta\mathcal{K}(t_1, t_2; t_3, t_4) \, = \, \int dt_5 dt_6 \, \frac{\delta^3 S_c[\Psi_0]}{\delta\Psi_0(t_1, t_2) \delta\Psi_0(t_3, t_4) \delta\Psi_0(t_5, t_6)} \, \Psi_1(t_5, t_6) \, .
	\label{delta Kernel}
	\end{equation}
Let us formally denote the $t_1$ - $t_4$ integrals in Eq.(\ref{S_2[epsilon]1}) by
	\begin{equation}
		\delta k_t \, \delta(t - t') \, = \, \int dt_1 dt_2 dt_3 dt_4 \, u_{0, t}(t_1, t_2) \, \delta\mathcal{K}(t_1, t_2; t_3, t_4) \, u_{0, t'}(t_3, t_4) \, ,
	\label{delta k}
	\end{equation}
because this is related to the eigenvalue shift due to $\delta \mathcal{K}$ up to normalization.
Then, we can write the quadratic action (\ref{S_2[epsilon]1}) as
	\begin{equation}
		S_2[\varepsilon] \, = \, - \, \frac{N}{4} \int dt \, \delta k_t \, \varepsilon^2(t) \, .
	\label{S_2[epsilon]2}
	\end{equation}

We now give a formal proof that the quadratic action (\ref{S_2[epsilon]2}) is equivalent to the quadratic action of Eq.(\ref{S[f]}).
This statement can be easily seen from the following identity:
	\begin{align}
		& \int dt_1 dt_2 dt_3 dt_4 \ u_{0, t}(t_1, t_2) \, \frac{\delta^3 S_{\rm c}[\Psi_0]}{\delta \Psi_0(t_1, t_2) \delta \Psi_0(t_3, t_4) \delta \Psi_0(t_5, t_6)} \, u_{0, t'}(t_3, t_4) \nonumber\\
		\, &= \, - \, \int dt_3 dt_4 \ \mathcal{K}(t_3, t_4; t_5, t_6) \ \frac{\delta^2 \Psi_{0, f}(t_3, t_4)}{\delta f(t) \delta f(t')} \bigg|_{f(t)=t} \ .
	\label{id}
	\end{align}
This identity is derived as follows.
In the zero mode equation $\int \mathcal{K} \cdot u_0=0$, rewriting the kernel as derivatives of $S_{\rm c}$ as in the first line of Eq.(\ref{kernel}),
and taking a derivative of this equation respect to $f(t')$, one finds
	\begin{align}
		0 \, &= \, \int dt_1 dt_2 dt_3 dt_4 \ \frac{\delta \Psi_{0, f}(t_1, t_2)}{\delta f(t)} \bigg|_{f(t)=t} \cdot
		\frac{\delta^3 S_{\rm c}[\Psi_{0,f}]}{\delta \Psi_{0,f}(t_1, t_2) \delta \Psi_{0,f}(t_3, t_4) \delta \Psi_{0,f}(t_5, t_6)} \cdot
		\frac{\delta \Psi_{0, f}(t_3, t_4)}{\delta f(t')} \bigg|_{f(t')=t'} \nonumber\\
		&\ + \, \int dt_3 dt_4 \ \frac{\delta^2 S_{\rm c}[\Psi_{0,f}]}{\delta \Psi_{0,f}(t_3, t_4) \delta \Psi_{0,f}(t_5, t_6)}
		\cdot \frac{\delta^2 \Psi_{0, f}(t_3, t_4)}{\delta f(t) \delta f(t')} \bigg|_{f(t)=t} \, ,
	\end{align}
where we used the zero mode expression (\ref{zero mode}).
Since $S_{\rm c}$ is invariant under the reparametrization, we can change the argument of $S_{\rm c}$ from $\Psi_{0, f}$ to $\Psi_0$. Then, we get the identity (\ref{id}).

We note that at next cubic level, one will have disagreement and the zero mode dynamics will not give the Schwarzian derivative.
This follows from the further identity:
	\begin{align}
		&\ \int dt_5 dt_6 \ \frac{\delta^2 S_{\rm c}[\Psi_0]}{\delta \Psi_0(t_5, t_6) \delta \Psi_0(t_7, t_8)} \cdot
		\frac{\delta^3 \Psi_{0, f}(t_5, t_6)}{\delta f(t) \delta f(t') \delta f(t'')} \bigg|_{f(t)=t} \nonumber\\
		&= \, - \, \int dt_1 dt_2 dt_3 dt_4 dt_5 dt_6 \ \frac{\delta^4 S_{\rm c}[\Psi_0]}{\delta \Psi_0(t_1, t_2) \delta \Psi_0(t_3, t_4) \delta \Psi_0(t_5, t_6) \delta \Psi_0(t_7, t_8)}
		\, u_{0, t}(t_1, t_2) \, u_{0, t'}(t_3, t_4) \, u_{0, t''}(t_5, t_6) \nonumber\\
		&- \, 3 \int dt_3 dt_4 dt_5 dt_6 \ \frac{\delta^3 S_{\rm c}[\Psi_0]}{\delta \Psi_0(t_3, t_4) \delta \Psi_0(t_5, t_6) \delta \Psi_0(t_7, t_8)}
		\, \frac{\delta^2 \Psi_{0, f}(t_3, t_4)}{\delta f(t) \delta f(t')} \bigg|_{f(t)=t} \, \frac{\delta \Psi_{0, f}(t_5, t_6)}{\delta f(t'')} \bigg|_{f(t)=t} \, ,
	\end{align}
where the second term in the right-hand side explains the expected discrepancy.

\section{Shift of the Classical Solution}
\label{sec:shift of the classical solution}
In large $N$ limit, the exact classical solution $\Psi_{\rm cl}$ is given by the solution of the saddle-point equation of the collective action (\ref{S_col}).
This classical solution corresponds to the one-point function:
	\begin{equation}
		 \big\langle \Psi(t_1, t_2) \big\rangle \, = \, \Psi_{\rm cl}(t_1, t_2) \, .
	\end{equation}
At the strict strong coupling limit, the classical solution is given by the critical solution $\Psi_0$, which is a solution of the saddle-point equation of the critical action (\ref{S_c}).
One can then develop a perturbative $1/J$ expansion for the full solution .

\subsection{Evaluation of $\Psi_1$}
\label{sec:evaluation of psi_1}
Let us consider the first order shift $\Psi_1$ of the classical solution from the critical solution induced by the breaking term.
We start with the naive delta function breaking term of the action $S_{\rm col}$ (\ref{S_col}).
Substitution of $\Psi_{\rm cl}=\Psi_0+\Psi_1$ gives
	\begin{equation}
		\int dt_3 dt_4 \, \mathcal{K}(t_1, t_2; t_3, t_4) \Psi_1(t_3, t_4) \, = \, \partial_1 \delta(t_{12}) \, ,
	\label{Psi_1-eq}
	\end{equation}
where the kernel is given in Eq.(\ref{kernel}).

It is useful to separate the $J$ dependence from the bi-local field by
	\begin{equation}
		\Psi_{\rm cl}(t_1, t_2) \, = \, J^{-\frac{2}{q}} \, \Psi_0(t_1, t_2) \, + \, \cdots \, ,
	\label{Psi_cl}
	\end{equation}
so that the critical solution $\Psi_0$, now reads
	\begin{equation}
		 \Psi_0(t_1, t_2) \, = \, b \ \frac{{\rm sgn}(t_{12})}{|t_{12}|^{\frac{2}{q}}} \, ,
	\end{equation}
with
	\begin{equation}
		 b \, = \, - \, \left[ \frac{\tan\left( \frac{\pi}{q} \right)}{2\pi} \left( 1- \frac{2}{q} \right) \right]^{\frac{1}{q}} \, .
	\label{define b}
	\end{equation}
Now the kernel (\ref{kernel}) does not have the explicit $J^2$ factor in the second term, and such rescaled kernel denoted by $\mathcal{K}$ will be used in the rest of the paper.
Since
	\begin{equation}
		 \Psi_0^{-1}(t_1, t_2) \, = \, - \, b^{q-1} \ \frac{{\rm sgn}(t_{12})}{|t_{12}|^{2-\frac{2}{q}}} \, ,
	\end{equation}
and the kernel has dimension $\mathcal{K} \sim |t|^{-4+4/q}$, from dimension analysis $\Psi_1$ would need to be the form of
	\begin{equation}
		\Psi_1(t_1, t_2) \, = \, A \ \frac{\, {\rm sgn}(t_{12}) \, }{|t_{12}|^{\frac{4}{q}}} \, ,
	\label{Psi_(1)}
	\end{equation}
where $A$ is a $t$-independent coefficient. In checking this ansatz we have the following integral in the first term of the LHS of Eq.(\ref{Psi_1-eq}) 
	\begin{equation}
		A b^{2q-2} \int dt_3 dt_4 \ \frac{{\rm sgn}(t_{13}) \, {\rm sgn}(t_{24}) \, {\rm sgn}(t_{34})}{|t_{13}|^{2-\frac{2}{q}} \, |t_{24}|^{2-\frac{2}{q}} \, |t_{34}|^{\frac{4}{q}}} \, .
	\end{equation}
This type of integral is already evaluated in Appendix A of \cite{Polchinski:2016xgd}.
In general, the result is 
	\begin{align}
		\int dt_3 dt_4 \ \frac{{\rm sgn}(t_{13}) \, {\rm sgn}(t_{24}) \, {\rm sgn}(t_{34})}{|t_{13}|^{2\Delta} \, |t_{24}|^{2\Delta} \, |t_{34}|^{2\alpha}}
		=&\, - \pi^2 \left[ \frac{\sin(2\pi \alpha)\, + \, 2\sin(2\pi (\alpha+\Delta))
		\, + \, \sin(2\pi (\alpha+2\Delta))}{\sin(2\pi \alpha)\sin(2\pi \Delta)\sin(2\pi (\alpha+\Delta))\sin(2\pi (\alpha+2\Delta))} \right] \nonumber\\
		&\qquad \times \frac{\Big[ \sin(2\pi \Delta) + \sin(2\pi (\alpha+\Delta)) \Big] \Gamma(1-2\Delta)}{\Gamma(2\alpha)\Gamma(2\Delta)\Gamma(3-2\alpha-4\Delta)}
		\frac{{\rm sgn}(t_{12})}{|t_{12}|^{2\alpha+4\Delta-2}} \, .
	\label{Polchinski-integrals}
	\end{align}
Our interest is $\Delta=1-1/q$. For this case, the result is inversely proportional to $\Gamma(4/q-2\alpha-1)$.
If we plug $\alpha=2/q$ into this equation, we can see that the Gamma function in the denominator gives infinity: $\Gamma(4/q-2\alpha-1)=\Gamma(-1)=\infty$, while other part is finite.
Therefore, the first term of the LHS of Eq.(\ref{Psi_1-eq}) vanishes.
The second term is trivial to evaluate; however the resulting form does not agree with the naive $\delta$-function source in RHS.
Hence, we conclude that the $\delta'$-source is only matched in the non-perturbative solution level, where all the $1/J$ corrections are summed over.

To proceed, consider a more general ansatz for $\Psi_1$:
	\begin{equation}
		\Psi_1(t_1, t_2) \, = \, B_1 \ \frac{\, {\rm sgn}(t_{12}) \, }{|t_{12}|^{\frac{2}{q}+2s}} \, ,
	\label{Psi_1 ansatz}
	\end{equation}
where $B_1$ is a $t$-independent coefficient.
The parameter $s$ has to be $s>0$, because the dimension of $\Psi_1$ needs to be less than the scaling dimension of $\Psi_0$.
Now using this ansatz, we are going to evaluate Eq.(\ref{Psi_1-eq}).
The integral of the first term of LHS of Eq.(\ref{Psi_1-eq}) is evaluated from Eq.(\ref{Polchinski-integrals}) with $\Delta=1-1/q$ and $\alpha=s+1/q$ as
	\begin{equation}
		\frac{B_1 \, b^{2q-2} \, \pi^2 \cot\left( \frac{\pi}{q} \right) \Gamma\left( \frac{2}{q} -1 \right)}
		{\sin\left(\pi \left( \frac{1}{q} + s \right)\right) \cos\left( \pi \left( s - \frac{1}{q} \right)\right)
		\Gamma\left(\frac{2}{q}+2s\right) \Gamma\left( 2-\frac{2}{q} \right) \Gamma\left( \frac{2}{q} - 2s -1 \right)} \, 
		\frac{{\rm sgn}(t_{12})}{|t_{12}|^{2-\frac{2}{q}+2s}} \, .
	\end{equation}
Hence, after a slight manipulation the LHS of Eq.(\ref{Psi_1-eq}) becomes
	\begin{equation}
		\int dt_3 dt_4 \, \mathcal{K}(t_1, t_2; t_3, t_4) \Psi_1(t_3, t_4) \, = \, (q-1) B_1 b^{q-2} \gamma(s, q) \, \frac{{\rm sgn}(t_{12})}{|t_{12}|^{2-\frac{2}{q}+2s}} \, ,
	\label{regularized source eq}
	\end{equation}
where we used Eq.(\ref{define b}) and we defined
	\begin{equation}
		\gamma(s, q) \, = \, 1 - \frac{\pi \, \Gamma\left( \frac{2}{q} \right)}{q \sin\left(\pi \left( \frac{1}{q} + s \right)\right) \cos\left( \pi \left( s - \frac{1}{q} \right)\right)
		\Gamma\left(\frac{2}{q}+2s\right) \Gamma\left( 3-\frac{2}{q} \right) \Gamma\left( \frac{2}{q} - 2s -1 \right)} \, .
	\label{gamma(s, q)}
	\end{equation}
Now we note that for $s=1/2$, $\gamma(1/2, q)=0$, so that the ansatz (\ref{Psi_1 ansatz}) would be the homogeneous equation associated with Eq.(\ref{Psi_1-eq}).
This limit $s \to 1/2$ therefore leads to the following first order shift of the background:
	\begin{equation}
		\Psi_{\rm cl}(t_1, t_2) \, = \, J^{-\frac{2}{q}} \Big[ \, \Psi_0(t_1, t_2) \, + \, J^{-1} \, \Psi_1(t_1, t_2) \, + \, \cdots \Big] \, , 
	\end{equation}
with
	\begin{equation}
		\Psi_0(t_1, t_2) \, = \, b \ \frac{\, {\rm sgn}(t_{12}) \, }{|t_{12}|^{\frac{2}{q}}} \, , \qquad \quad
		\Psi_1(t_1, t_2) \, = \, B_1 \ \frac{\, {\rm sgn}(t_{12}) \, }{|t_{12}|^{\frac{2}{q}+1}} \, .
	\label{Psi_1}
	\end{equation}

We will however keep the parameter $s$ infinitesimally away from $1/2$ as a regularization.
Then, 
	\begin{align}
		\gamma(s, q) \, = \, \frac{6 \, q \, \gamma}{(q-1)\, b^{q-1}} \big( s - \tfrac{1}{2} \big) \, + \, \mathcal{O}\Big((s-\tfrac{1}{2})^2 \Big) \, ,
		\label{gamma(s,q)-expansion}
	\end{align}
where $\gamma$ is defined in Eq.(\ref{gamma}), 
and the RHS in Eq.(\ref{regularized source eq}) can be interpreted as a regularized non-zero source term of the form 
	\begin{equation}
		Q_s(t_1, t_2) \, \equiv \, (s- \tfrac{1}{2}) \, 6q B_1 b^{-1} \gamma \ \frac{{\rm sgn}(t_{12})}{|t_{12}|^{2-\frac{2}{q}+2s}} \, + \, \mathcal{O}\big( (s-\tfrac{1}{2})^2 \big) \, .
	\label{Q_s}
	\end{equation}
The $\gamma$ is obtained by expanding $\gamma(s, q)$ (\ref{gamma(s, q)}) around $s=1/2$ so that 
	\begin{equation}
		\gamma \, = \, \frac{(q-1)b^{q-1}}{6q} \ \gamma'(s=\tfrac{1}{2}, q) \, .
	\end{equation}
Here, the prime denotes a derivative respect to $s$.
We use this regularized source to define the regularized breaking term by
	\begin{equation}
		\int \, \big[ \Psi_{ f} \big]_s \, \equiv \, - \, \lim_{s \to \frac{1}{2}} \int dt_1 dt_2 \, \Psi_{ f}(t_1, t_2) \, Q_s(t_1, t_2) \, .
	\end{equation}

Finally, the coefficient $B_1$ can be deduced from the numerical result found in \cite{Maldacena:2016hyu}.
Comparison of the two results gives the relation:
	\begin{equation}
		\frac{B_1}{bJ} \,=\, \frac{\alpha_G}{\mathcal{J}} \, ,
	\label{alpha_G}
	\end{equation}
with the numerical approximated value of $\alpha_G$ established in \cite{Maldacena:2016hyu} 
	\begin{equation}
		\alpha_G \, \approx \, \frac{2(q-2)}{16/\pi+6.18(q-2)+(q-2)^2} \, ,
	\end{equation}
and $\mathcal{J}=\frac{\sqrt{q}}{2^{\frac{q-1}{2}}}J$.

\subsection{Evaluation of $\Psi_2$}
\label{sec:evaluation of psi_2}
Now we would like to go further higher order term in the expansion of the classical solution.
This term is given by
	\begin{equation}
		\Psi_{\rm cl}(t_1, t_2) \, = \, J^{-\frac{2}{q}} \Big[ \, \Psi_0(t_1, t_2) \, + \, J^{-1} \, \Psi_1(t_1, t_2) \, + \, J^{-2} \, \Psi_2(t_1, t_2) \, + \, \cdots \Big] \, , 
	\label{Psi expansion for Psi_2}
	\end{equation}
with
	\begin{equation}
		\Psi_2(t_1, t_2) \, = \, B_2 \ \frac{\, {\rm sgn}(t_{12}) \, }{|t_{12}|^{\frac{2}{q}+2}} \, ,
	\end{equation}
where $B_2$ is a $t$-independent coefficient.
The dimension of $\Psi_2$ is already fixed by $\Psi_1$, so what we need to do is just to fix the coefficient $B_2$.
Substituting the above expansion of the classical field into the critical action $S_{\rm c}$ (\ref{S_c}) and expanding it,
one finds that the equation determining $\Psi_2$ is given by
	\begin{align}
		&\ \int dt_3 dt_4 \, \mathcal{K}(t_1, t_2; t_3, t_4) \Psi_2(t_3, t_4) \nonumber\\
		\, =& \ - \, [\Psi_0^{-1} \star \Psi_1 \star \Psi_0^{-1} \star \Psi_1 \star \Psi_0^{-1}](t_1, t_2)
		\, - \, \frac{(q-1)(q-2)}{2} \ \Psi_0^{q-3}(t_1, t_2) \, \Psi_1^2(t_1, t_2) \, ,
	\label{Psi_2-eq}
	\end{align}
where the star product is defined by $[A \star B](t_1, t_2) \equiv \int dt_3 \, A(t_1, t_3) B(t_3 , t_2)$.
Now, we are going to evaluate each term of this equation.
For the first term in the LHS is again given by Eq.(\ref{Polchinski-integrals}) with $\Delta=1-1/q$ and $\alpha=1/q+1$ as
	\begin{equation}
		({\rm LHS\ 1st}) \, = \, 2\pi \, B_2 \, b^{2q-2} \ \frac{q(q-1)(3q-2)}{(q^2-4)\tan(\frac{\pi}{q})} \ \frac{{\rm sgn}(t_{12})}{|t_{12}|^{4-\frac{2}{q}}} \, .
	\end{equation}
For the first term of the RHS, we need to use Eq.(\ref{Polchinski-integrals}) twice.
First for the middle of the term: $\Psi_1\star\Psi_0^{-1}\star\Psi_1$, and then for the result sandwiched by the remaining $\Psi_0^{-1}$'s. Then, we have
	\begin{equation}
		({\rm RHS\ 1st}) \, = \, - \, B_1^2 \, b^{3(q-1)} \, \frac{2\pi^2 q^2 (q-1)(3q-2)}{(q-2)^2} \ \frac{{\rm sgn}(t_{12})}{|t_{12}|^{4-\frac{2}{q}}} \, .
	\end{equation}
The second terms in the LHS and RHS are trivially evaluated.
Therefore, now one can see that all terms have the same $t_{12}$ dependence.
Then, comparing their coefficients, we finally fix $B_2$ as
	\begin{equation}
		B_2 \, = \, - \, \frac{B_1^2}{b} \left( \frac{q+2}{8q} \right) \left[ (q-2) + (3q-2) \tan^2\left( \frac{\pi}{q} \right) \right] \, .
	\end{equation}

\subsection{All Order Evaluation in $q>2$}
\label{sec:all order evaluation in q>2}
In this subsection, we extend our previous perturbative expansion of the classical solution to all order contributions in the $1/J$ expansion.
Because of the dimension of $\Psi_1$ (\ref{Psi_1}), the time-dependence is already fixed for all order as in Eq.(\ref{Psi_n}).
Therefore, we only need to determine the coefficient $B_n$, and in this subsection we will give a recursion relation which fixes the coefficients.
However, we will not use this subsection's result in the rest of the paper,
so readers who are interested only in the first few terms in the $1/J$ expansion (\ref{Psi expansion for Psi_2}) may skip this subsection and move on to Section \ref{sec:two-point function}.
As we saw in Section \ref{sec:evaluation of psi_1}, the structure of the classical solution in $q=2$ model is different from $q>2$ case.
In this subsection, we focus on $q>2$ case.

We generalize the expansion (\ref{Psi expansion for Psi_2}) to all order by
	\begin{equation}
		\Psi_{\rm cl}(t_1, t_2) \, = \, J^{-\frac{2}{q}} \, \sum_{m=0}^{\infty} J^{-m} \, \Psi_m(t_1, t_2) \, .
	\end{equation}
Now, we substitute this expansion into the critical action $S_{\rm c}$ (\ref{S_c}).
As we saw before, the kinetic term does not contribute to the perturbative analysis when $q>2$; therefore, we discard the kinetic term here.
The contribution of the kinetic term will be recovered in the full classical solution with correct UV boundary conditions.
Hence, the saddle-point equation is now formally written as
	\begin{equation}
		0 \ = \ \left[ \sum_{m=0}^{\infty} J^{-m} \, \Psi_m(t_1, t_2) \right]^{-1} \ + \ \left[ \sum_{m=0}^{\infty} J^{-m} \, \Psi_m(t_1, t_2) \right]^{q-1} \, .
	\label{saddle-eq all order}
	\end{equation}
Using the multinomial theorem, each term can be reduced to polynomials of $\Psi_m$'s.
Substituting these results into Eq.(\ref{saddle-eq all order}) leads the saddle-point equation written in terms polynomials with all order of $1/J$ expansion.
From this equation, one can further pick up order $\mathcal{O}(J^{-n})$ terms.
For $n=0$, it is the equation of $\Psi_0$.
Therefore, we consider $n\ge1$ case, which is given by
	\begin{align}
		0 \, &= \, \sum_{k_1 + 2k_2 + \cdots=n} \ (-1)^{k_1+k_2+\cdots} \ \frac{(k_1 + k_2 + \cdots)!}{k_1!k_2!k_3! \cdots}
		\times \left[ \Psi_0^{-1} \star \Big( \Psi_1 \star \Psi_0^{-1} \Big)^{k_1} \star \Big( \Psi_2 \star \Psi_0^{-1} \Big)^{k_2} \star \cdots \right](t_1, t_2) \nonumber\\
		&\quad + \, \sum_{k_1 + 2k_2 + \cdots=n} \ \frac{(q-1)!}{k_0!k_1!k_2! \cdots}
		\, \times \, \Psi_0^{k_0}(t_1, t_2) \, \Psi_1^{k_1}(t_1, t_2) \, \Psi_2^{k_2}(t_1, t_2) \cdots \, ,
	\end{align}
with $k_0=q-(1+k_1+\cdots+k_{n-1})$.
Let us consider this order $\mathcal{O}(J^{-n})$ equation more.
Because of the constraint $k_1 + 2k_2 + \cdots=n$, we know that $k_{n+1}=k_{n+2}= \cdots = 0$.
Also the same constraint implies that $k_n=0$ or $1$, and when $k_n=1$, then $k_1 = k_2 = \cdots = k_{n-1} =0$.
Therefore, it is useful to separate $k_n=1$ terms from $k_n=0$ ones.
After this separation, the order $\mathcal{O}(J^{-n})$ equation is reduced to a more familiar form:
	\begin{align}
		&\quad \int dt_3 dt_4 \, \mathcal{K}(t_1, t_2; t_3, t_4) \Psi_n(t_3, t_4) \nonumber\\
		&= \, - \sum_{k_1 + 2k_2 + \cdots + (n-1)k_{n-1} =n} (-1)^{k_1+\cdots+k_{n-1}} \ \frac{(k_1 + \cdots + k_{n-1})!}{k_1! \cdots k_{n-1}!} \nonumber\\
		&\hspace{140pt} \times \left[ \Psi_0^{-1} \star \Big( \Psi_1 \star \Psi_0^{-1} \Big)^{k_1} \star \cdots \star \Big( \Psi_{n-1} \star \Psi_0^{-1} \Big)^{k_{n-1}} \right](t_1, t_2) \nonumber\\
		&\quad - \, \sum_{k_1 + 2k_2 + \cdots + (n-1)k_{n-1}=n} \ \frac{(q-1)!}{k_0!k_1! \cdots k_{n-1}!}
		\, \times \, \Psi_0^{k_0}(t_1, t_2) \, \Psi_1^{k_1}(t_1, t_2) \, \cdots \, \Psi_{n-1}^{k_{n-1}}(t_1, t_2) \, ,
	\label{Psi_n-eq}
	\end{align}
where $k_0=q-(1+k_1+\cdots+k_{n-1})$.
This is the equation which determines $\Psi_n$ from $\{\Psi_0, \Psi_1, \cdots, \Psi_{n-1}\}$ sources.
However, we already know the $t_{12}$ dependence of $\Psi_n(t_1, t_2)$.
Namely,
	\begin{equation}
		\Psi_n(t_1, t_2) \, = \, B_n \ \frac{{\rm sgn}(t_{12})}{|t_{12}|^{\frac{2}{q}+n}} \, .
	\label{Psi_n}
	\end{equation}
Therefore, we only need to determine the coefficient $B_n$.
Probably it is hard to evaluate the star products in the RHS of Eq.(\ref{Psi_n-eq}) by direct integrations of $t$'s, and it is better to use momentum space representations.
	\begin{equation}
		\Psi_m(t_1, t_2) \, = \, B_m \ \int \frac{d\omega}{2\pi} \, e^{-i\omega t_{12}} \, \Psi_m(\omega) \, ,
	\label{Fourier}
	\end{equation}
where we excluded the coefficient $B_m$ from $\Psi_m(\omega)$ for later convenience, and $\Psi_m(\omega) = C_m \ |\omega|^{\frac{2}{q}+m-1} \, {\rm sgn}(\omega)$, 
with
	\begin{equation}
		C_m \, \equiv \, i \, 2^{1-m-\frac{2}{q}} \sqrt{\pi} \ \frac{\Gamma(1-\frac{1}{q}-\frac{m}{2})}{\Gamma(\frac{1}{q}+\frac{m}{2}+\frac{1}{2})} \, .
	\label{C_m}
	\end{equation}
With this definition of $C_m$, we can write the inverse of the critical solution as
	\begin{equation}
		\Psi_0^{-1}(t_1, t_2) \, = \, \int \frac{d\omega}{2\pi} \, e^{-i\omega t_{12}} \, \Psi_0^{-1}(\omega)
		\, = \, - \, b^{q-1} \, C_{2-\frac{4}{q}} \, \int \frac{d\omega}{2\pi} \, e^{-i\omega t_{12}} \, |\omega|^{1-\frac{2}{q}} \, {\rm sgn}(\omega) \, .
	\end{equation}
Now, we can evaluate each term in Eq.(\ref{Psi_n-eq}) using these Fourier transforms.
Then, every term has the same $\omega$ integral; therefore, comparing the coefficients, one obtains
	\begin{align}
		&\quad b^{q-2} \Big[ \, (q-1) \, C_{2+n-\frac{4}{q}} \, - \, b^q \, C_{2-\frac{4}{q}}^2 \, C_n \, \Big] B_n \nonumber\\
		&= \, - \sum_{k_1 + 2k_2 + \cdots + (n-1)k_{n-1} =n} (-1)^{k_1+\cdots+k_{n-1}} \ \frac{(k_1+\cdots+k_{n-1})!}{k_1! \cdots k_{n-1}!} \nonumber\\
		&\hspace{160pt} \times \Big( - b^{q-1} \, C_{2-\frac{4}{q}} \Big)^{k_1+\cdots+k_{n-1}+1} \Big( B_1 \, C_1 \Big)^{k_1} \cdots \Big( B_{n-1} \, C_{n-1} \Big)^{k_{n-1}} \nonumber\\
		&\quad - \, \sum_{k_1 + 2k_2 + \cdots + (n-1)k_{n-1}=n} \ \frac{(q-1)!}{k_0!k_1! \cdots k_{n-1}!}
		\, \times \, b^{k_0} B_1^{k_1} \cdots B_{n-1}^{k_{n-1}} \, C_{2+n-\frac{4}{q}} \, ,
	\label{recursion relation}
	\end{align}
with $k_0=q-(1+k_1+\cdots+k_{n-1})$.
This is the recursion relation which determines $B_n$ from $\{B_1, B_{2}, \cdots, B_{n-1}\}$.
Note that $C_m$'s are a priori known numbers as defined in Eq.(\ref{C_m}).

\section{Two-point Function}
\label{sec:two-point function}
In this section, we consider the bi-local two-point function:
	\begin{equation}
		\Big\langle \Psi(t_1, t_2) \Psi(t_3, t_4) \Big\rangle \, ,
	\end{equation}
where the expectation value is evaluated by the path integral (\ref{eq:collective partition function}).
After the Faddeev-Popov prosedure and changing the integration variable as we discussed in Section \ref{sec:intro}, this two-point function becomes
	\begin{equation}
		\Big\langle \Psi_f(t_1, t_2) \Psi_f(t_3, t_4) \Big\rangle \, ,
	\end{equation}
where now the expectation value is evaluated by the gauged path integral (\ref{eq:gauged collective partition function}).

Now, we expand the bi-local field around the shifted background classical solution $\Psi_{\rm cl}=\Psi_0+J^{-1} \Psi_1$.
Namely,
	\begin{equation}
		\Psi(t_1, t_2) \, = \, \Psi_0(t_1, t_2) \, + \, \frac{1}{J} \, \Psi_1(t_1, t_2) \, + \, \sqrt{\frac{2}{N}} \ \overline{\eta}(t_1, t_2) \, ,
	\end{equation}
where we have rescaled the entire field $\Psi$ by $J^{2/q}$, and $\overline{\eta}$ is a quantum fluctuation, but the zero mode is eliminated from its Hilbert space.
Therefore, the two-point function is now decomposed as 
	\begin{equation}
		\Big\langle \Psi_f(t_1, t_2) \Psi_f(t_3, t_4) \Big\rangle
		\, = \, \Big\langle \Psi_{{\rm cl},f}(t_1, t_2) \Psi_{{\rm cl},f}(t_3, t_4) \Big\rangle \, + \, \frac{2}{N} \, \Big\langle \overline{\eta}(t_1, t_2) \overline{\eta}(t_3, t_4) \Big\rangle \, .
	\label{two-point decomposition}
	\end{equation}
The second term in the RHS is the bi-local propagator $\mathcal{D}$ determined by Eq.(\ref{Green's eq}),
which was already evaluated in I for $q=4$ (and also in \cite{Polchinski:2016xgd,Maldacena:2016hyu}) as
	\begin{align}
		\mathcal{D}(t_1, t_2; t_3, t_4) \, = \, - \, {\rm sgn}(t_- t_-') \, \frac{8}{N\sqrt{\pi}} \sum_{m=1}^{\infty} &\int d\omega \, 
		\frac{e^{-i\omega(t_+-t_+')}}{\sin(\pi p_m)} \frac{p_m^2}{p_m^2+(3/2)^2} \nonumber\\
		&\times \left[ J_{-p_m}(|\omega t_-|) \, + \, \frac{p_m+\frac{3}{2}}{p_m-\frac{3}{2}} \, J_{p_m}(|\omega t_-|) \right] J_{p_m}(|\omega t_-'|) \, ,
	\end{align}
where $p_m$ are the solutions of $2p_m/3=-\tan(\pi p_m/2)$, and $t_{\pm}=(t_1\pm t_2)/2$ and $t_{\pm}'=(t_3\pm t_4)/2$.

Therefore, in this section let us focus on the first term in the RHS of Eq.(\ref{two-point decomposition}).
Expanding the classical field up to the second order, one has
	\begin{align}
		&\quad\ \, \Big\langle \Psi_{{\rm cl},f}(t_1, t_2) \Psi_{{\rm cl},f}(t_3, t_4) \Big\rangle \nonumber\\
		&= \, \Big\langle \Psi_{0,f}(t_1, t_2) \Psi_{0,f}(t_3, t_4) \Big\rangle
		\, + \, \frac{1}{J} \left[ \Big\langle \Psi_{0,f}(t_1, t_2) \Psi_{1,f}(t_3, t_4) \Big\rangle \, + \, \binom{t_1 \leftrightarrow t_3}{t_2 \leftrightarrow t_4} \right] \, + \, \cdots \, ,
	\end{align}
where
	\begin{align}
		\Psi_{0,f}(t_1, t_2) \, &= \, \Big| f'(t_1) f'(t_2) \Big|^{\frac{1}{q}} \ \Psi_0(f(t_1), f(t_2)) \, , \nonumber\\
		\Psi_{1,f}(t_1, t_2) \, &= \, \Big| f'(t_1) f'(t_2) \Big|^{\frac{1}{q}+\frac{1}{2}} \ \Psi_1(f(t_1), f(t_2)) \, .
	\end{align}
Now, we consider an infinitesimal reparametrization $f(t)=t+\varepsilon(t)$.
Then, the classical fields are expanded as
	\begin{align}
		\Psi_{0,f}(t_1, t_2) \, &= \, \Psi_0(t_1, t_2) \, + \, \int dt \, \varepsilon(t) \, u_{0,t}(t_1, t_2) \, + \, \cdots \, , \nonumber\\
		\Psi_{1,f}(t_1, t_2) \, &= \, \Psi_1(t_1, t_2) \, + \, \int dt \, \varepsilon(t) \, u_{1,t}(t_1, t_2) \, + \, \cdots \, ,
	\end{align}
where
	\begin{equation}
		u_{0,t}(t_1, t_2) \, \equiv \, \frac{\partial \Psi_{0, f}(t_1, t_2)}{\partial f(t)} \bigg|_{f(t)=t} \, , \qquad
		u_{1,t}(t_1, t_2) \, \equiv \, \frac{\partial \Psi_{1, f}(t_1, t_2)}{\partial f(t)} \bigg|_{f(t)=t} \, .
	\end{equation}
Therefore, in the quadratic order of $\varepsilon$, the classical field two-point function is now written in term of the two-point function of $\varepsilon$.
For later convenience, it is better to write down this as momentum space integral as
	\begin{align}
		&\qquad \Big\langle \Psi_{{\rm cl},f}(t_1, t_2) \Psi_{{\rm cl},f}(t_3, t_4) \Big\rangle \nonumber\\
		\, &= \, \int \frac{d\omega}{2\pi} \, \langle \varepsilon(\omega) \varepsilon(-\omega) \rangle \bigg[ \, u_{0,\omega}^*(t_1, t_2) u_{0,\omega}(t_3, t_4)
		\, + \, \frac{1}{J} \left( u_{0,\omega}^*(t_1, t_2) u_{1,\omega}(t_3, t_4) \, + \, \binom{t_1 \leftrightarrow t_3}{t_2 \leftrightarrow t_4} \right) \, + \, \cdots \bigg] \, .
	\end{align}
Let us first evaluate the $\varepsilon$ two-point function.
The collective coordinate action is given in Eq.(\ref{S[f]}).
Expanding $f(t)=t+\varepsilon(t)$, the quadratic action of $\varepsilon$ can be obtained from this action.
Hence, the two-point function in momentum space is 
	\begin{equation}
		\langle \varepsilon(\omega) \varepsilon(-\omega) \rangle \, = \, \frac{24\pi J}{\alpha N} \, \frac{1}{\omega^4} \, .
	\end{equation}
One can also Fourier transform back to the time representation to get 
	\begin{equation}
		\langle \varepsilon(t_1) \varepsilon(t_2) \rangle \, = \, \frac{2\pi J}{\alpha N} \ |t_{12}|^3 \, .
	\end{equation}
Next, we evaluate $u_0$ and $u_1$.
Taking the derivative respect to $f(t)$, one obtains
	\begin{align}
		u_{0,t}(t_1, t_2) \, &= \, \frac{1}{q} \left[ \, \delta'(t_1-t) \, + \, \delta'(t_2-t) \, - \, 2 \left( \frac{\delta(t_1-t) - \delta(t_2-t)}{t_1-t_2} \right) \, \right] \, \Psi_0(t_1, t_2) \, , \nonumber\\
		u_{1,t}(t_1, t_2) \, &= \, \frac{2+q}{2q} \left[ \, \delta'(t_1-t) \, + \, \delta'(t_2-t) \, - \, 2 \left( \frac{\delta(t_1-t) - \delta(t_2-t)}{t_1-t_2} \right) \, \right] \, \Psi_1(t_1, t_2) \nonumber\\
		&= \, \frac{(2+q)B_1}{2b} \, \frac{u_{0, t}(t_1, t_2)}{|t_{12}|} \, .
	\label{u_{0,t}}
	\end{align}
After some manipulation, one can show that the momentum space expressions are given by 
	\begin{align}
		u_{0,\omega}(t_1, t_2) \, &= \, - \, \frac{ib\sqrt{\pi}}{q} \, \frac{|\omega|^{\frac{3}{2}} \, {\rm sgn}(\omega t_-)}{|2 \, t_-|^{\frac{2}{q}-\frac{1}{2}}} \
		e^{i\omega t_+} \, J_{\frac{3}{2}}(|\omega t_-|) \, , \nonumber\\
		u_{1,\omega}(t_1, t_2) \, &= \, \frac{(2+q)B_1}{4b} \, \frac{u_{0, \omega}(t_1, t_2)}{|t_-|} \, .
	\label{u_{0,omega}}
	\end{align}
Using the two-point function of $\varepsilon$ and above $u_0$ and $u_1$ expressions, finally the two-point function (\ref{two-point decomposition}) up to order $J^0$ is given by
	\begin{align}
		\Big\langle \Psi_f(t_1, t_2) \Psi_f(t_3, t_4) \Big\rangle
		\, &= \, \frac{12}{\alpha N} \, \bigg[ \, J \, + \, \frac{(2+q)B_1}{4b} \left( \frac{1}{|t_-|} + \frac{1}{|t_-'|} \right) \bigg]
		\int \frac{d\omega}{\omega^4} \, u_{0, \omega}^*(t_1, t_2) u_{0, \omega}(t_3, t_4) \nonumber\\
		&\qquad + \, \mathcal{D}(t_1, t_2; t_3, t_4) \, .
	\end{align}

What we have established therefore is the following.
What one has is first the leading ``classical" contribution to the bi-local two-point function which usually factorizes, due to the dynamics of the reparametrization symmetry mode.
It now represents the leading `big' contribution, as in \cite{Maldacena:2016hyu}, and a sub-leading one.
This is followed by the matter fluctuations given by the zero mode projected propagator of I \cite{Jevicki:2016bwu}.

\section{Finite Temperature}
\label{sec:finite temperature}
Up to here, we have been considering only zero-temperature solutions in the SYK model.
In this section, we will consider the finite-temperature solutions $\Psi_{1, \beta}$ and $\Psi_{2, \beta}$ and the tree-level free energy in the low temperature region.

\subsection{Classical Solutions}
\label{sec:classical solutions}
As we saw in Section \ref{sec:shift of the classical solution}, the $1/J$ expansion of the classical solution in the strongly coupling region is given by
	\begin{equation}
		\Psi_{\rm cl}(t_1, t_2) \, = \, J^{-\frac{2}{q}} \Big[ \, \Psi_0(t_1, t_2) \, + \, J^{-1} \, \Psi_1(t_1, t_2) \, + \, J^{-2} \, \Psi_2(t_1, t_2) \, + \, \cdots \Big] \, , 
	\label{Psi_cl expansion}
	\end{equation}
where
	\begin{equation}
		\Psi_0(t_1, t_2) \, = \, b \ \frac{\, {\rm sgn}(t_{12}) \, }{|t_{12}|^{\frac{2}{q}}} \, , \quad
		\Psi_1(t_1, t_2) \, = \, B_1 \ \frac{\, {\rm sgn}(t_{12}) \, }{|t_{12}|^{\frac{2}{q}+1}} \, , \quad
		\Psi_2(t_1, t_2) \, = \, B_2 \ \frac{\, {\rm sgn}(t_{12}) \, }{|t_{12}|^{\frac{2}{q}+2}} \, .
	\label{Psi_0,1,2}
	\end{equation}
In order to evaluate tree-level free energy, we first need finite-temperature versions of these classical solutions.
$\Psi_0$ is the solution of the strict strong coupling limit,
where the model exhibits an emergent conformal reparametrization symmetry: $t \to f(t)$ with the $\Psi_0$ transformation (\ref{reparametrization}).
Therefore, to obtain the finite-temperature version of $\Psi_0$, we just need to use $f(t)=\frac{\beta}{\pi}\tan(\frac{\pi t}{\beta})$ with the above transformation \cite{Kitaev:2015}.
This map maps the infinitely long zero-temperature time to periodic thermal circle. Thus, this gives us 
	\begin{equation}
		\Psi_{0, \beta}(t_1, t_2) \, = \, b \ \left[ \frac{\pi}{\beta \sin(\frac{\pi t_{12}}{\beta})} \right]^{\frac{2}{q}} \, {\rm sgn}(t_{12}) \, .
	\label{Psi_0-beta}
	\end{equation}

Since $\Psi_1$ and $\Psi_2$ are the shifts of the classical solution from the strict IR limit, they do not enjoy the reparametrization symmetry.
Therefore, we cannot use the above method to get their finite-temperature counterparts.
However, we can approximate finite-temperature solutions by mapping the zero-temperature solutions onto a thermal circle and summing over all image charges:
	\begin{equation}
		\Psi_{\beta}(t_{12}) \, = \, \sum_{m=-\infty}^{\infty} \, (-1)^{m} \, \Psi_{\beta=\infty}(t_{12}+\beta m) \, .
	\end{equation}
In this approximation, the finite-temperature solutions (two-point function in terms of the fundamental fermions) trivially satisfy the KMS condition.
This approximation also works order by order in the $1/J$ expansion.
Therefore, after separating positive $m$ and negative $m$ and changing the labeling, one finds
	\begin{equation}
		\Psi_{1,\beta}(t_{12})
		\, = \, B_1 \left[ \ \sum_{m=0}^{\infty} \, \frac{(-1)^{m}}{(\beta m+t_{12})^{\frac{2}{q}+1}} \, - \, \sum_{m=1}^{\infty} \, \frac{(-1)^{m}}{(\beta m-t_{12})^{\frac{2}{q}+1}} \ \right] \, .
	\label{Psi_1-beta}
	\end{equation}
The summations of $m$ can be evaluated to give the Hurwitz zeta functions.
In the same way, we can approximate $\Psi_2$ in terms the Hurwitz zeta functions.

\begin{figure}[t!]
\vspace{0pt}
	\begin{center}
		\hspace{60pt} \scalebox{0.8}{\includegraphics{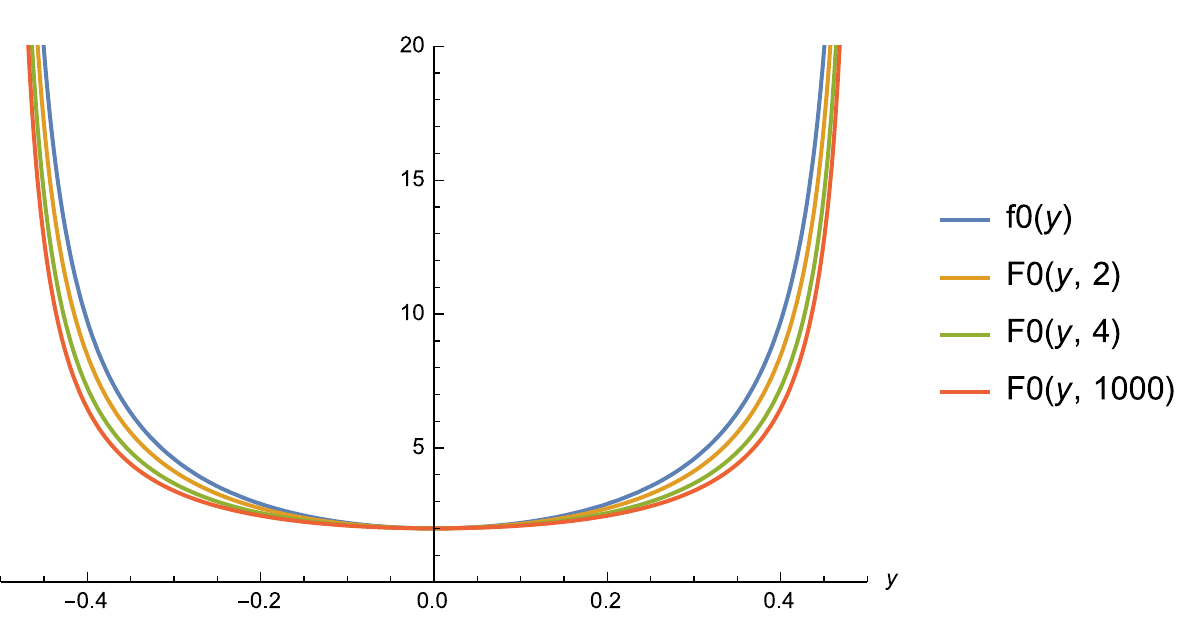}}
	\end{center}
	\caption{$f_0(y)$ and $F_0(y, q)$ with $q=2, 4, 1000$ in the range of $-\frac{1}{2} \le y \le \frac{1}{2}$.}
	\label{fig:f0andF0}
\end{figure}

In \cite{Maldacena:2016hyu}, Maldacena and Stanford obtained a first order shift of the classical solution in finite-temperature
through a numerical solution of the exact Schwinger-Dyson equation.
The above 'image charge' estimate $\Psi_{1,\beta}$ can be seen to agree well with the numerical ansatz.
The solution of \cite{Maldacena:2016hyu} is shown in their Eq.(3.122) reading:
	\begin{equation}
		\frac{\delta G(t_1, t_2)}{G_c(t_1, t_2)} \, = \, - \frac{\alpha_G}{\beta \mathcal{J}} \, f_0(t_{12}) \, , \qquad
		f_0(t_{12}) \, = \, 2 \, + \, \frac{\pi - \frac{2\pi|t_{12}|}{\beta}}{\tan|\frac{\pi t_{12}}{\beta}|} \, .
	\label{M/S solution}
	\end{equation}
with the notation, $G_c = \Psi_{0,\beta}$ and $\delta G = \Psi_{1,\beta}$.
We can see in Figure \ref{fig:f0andF0} that our approximated solution for $\Psi_{1,\beta}$ is pretty close to this solution.
It is more convenient to introduce a new variable
	\begin{equation}
		y \, \equiv \, \frac{|t_{12}|}{\beta} \, - \, \frac{1}{2} \, . \qquad \left( -\frac{1}{2} \le y \le \frac{1}{2} \right)
	\end{equation}
Then, we have $f_0(y) = 2 + 2\pi y \tan(\pi y)$.
On the other hand for the figure, we rewrite our approximated solution by
	\begin{equation}
		\frac{\Psi_{1,\beta}(t_{12})}{\Psi_{0, \beta}(t_{12})} \, = \, \frac{B_1}{2(2\pi)^{\frac{2}{q}}b\beta} \,
		\bigg[ \zeta\left( \tfrac{2}{q}+1, \tfrac{1}{4} \right) \, - \, \zeta\left( \tfrac{2}{q}+1, \tfrac{3}{4} \right) \bigg] \, \times F_0(y, q) \, ,
	\end{equation}
where 
	\begin{equation}
		F_0(y, q) \, \equiv \, (\cos \pi y)^{\frac{2}{q}} \, 
		\left[ \frac{\zeta\left( \tfrac{2}{q}+1, \tfrac{1}{4}+\tfrac{y}{2} \right) \, + \, \zeta\left( \tfrac{2}{q}+1, \tfrac{1}{4}-\tfrac{y}{2} \right) 
		\, - \, \zeta\left( \tfrac{2}{q}+1, \tfrac{3}{4}+\tfrac{y}{2} \right) \, - \, \zeta\left( \tfrac{2}{q}+1, \tfrac{3}{4}-\tfrac{y}{2} \right)}
		{\zeta\left( \tfrac{2}{q}+1, \tfrac{1}{4} \right) \, - \, \zeta\left( \tfrac{2}{q}+1, \tfrac{3}{4} \right)} \right] \, .
	\end{equation}
Here, we adjusted the normalization of $F_0$ so that $F_0(y=0, q)=2=f_0(y=0)$.
In Figure \ref{fig:f0andF0}, we plotted $f_0(y)$ and $F_0(y,q)$ with $q=2,4,1000$.
We can see that for any value of $q$, $F_0$ is pretty close to $f_0$ in all range of $y$.

We will now develop a small temperature expansion which will give further useful information about the finite temperature solution and also the free energy.
For this one expands the equation iteratively starting from $\Psi_{0, \beta}$ as sources.
We develop this method for $\Psi_{1,\beta}$ in the rest of this subsection.
The expansion of $\Psi_{0, \beta}$ solution (\ref{Psi_0-beta}) in the small temperature region is given by
	\begin{equation}
		\Psi_{0, \beta}(t_{12}) \, = \, b \ \frac{{\rm sgn}(t_{12})}{|t_{12}|^{\frac{2}{q}}}
		\left[ 1 + \frac{\pi^2}{3q} \left| \frac{t_{12}}{\beta} \right|^2 + \frac{(q+5)\pi^4}{90q^2} \left| \frac{t_{12}}{\beta} \right|^4 + \cdots \right] \, .
	\end{equation}
We then expand the finite-temperature solution $\Psi_{1, \beta}$ by 
	\begin{equation}
		\Psi_{1,\beta}(t_1, t_2) \, = \, B_1 \ \frac{{\rm sgn}(t_{12})}{|t_{12}|^{\frac{2}{q}+1}}
		\left[ 1 + c_{1,1} \left| \frac{t_{12}}{\beta} \right| + c_{1,2} \left| \frac{t_{12}}{\beta} \right|^2 + c_{1,3} \left| \frac{t_{12}}{\beta} \right|^3 + \cdots \right] \, ,
	\end{equation}
and then, using the equation of motion for $\Psi_1$ (\ref{Psi_1-eq}) we iteratively determine the coefficients $c_{1, i}$ starting from the lower order ones.
As we will see in the next subsection, to evaluate its free energy contribution, we need $a_1\equiv c_{1,3}$.
First we consider $\mathcal{O}(\beta^{-1})$ order. The equation in this order reads
	\begin{equation}
		B_1 c_{1,1} \beta^{-1} \int dt_3 dt_4 \, \mathcal{K}(t_1, t_2; t_3, t_4) \, \frac{{\rm sgn}(t_{34})}{|t_{34}|^{\frac{2}{q}}} \, = \, 0 \, ,
	\end{equation}
where $\mathcal{K}$ denotes the zero temperature kernel.
Using the formula in Eq.(\ref{Polchinski-integrals}), one can evaluate the left-hand side integrals.
In general, the integral does not vanish. Therefore, to satisfy the equation, we need $c_{1,1}=0$.
Next for $\mathcal{O}(\beta^{-2})$ order, we have an equation 
	\begin{align}
		&\ B_1 c_{1,2} \beta^{-2} \int dt_3 dt_4 \, \mathcal{K}(t_1, t_2; t_3, t_4) \frac{{\rm sgn}(t_{34})}{|t_{34}|^{\frac{2}{q}-1}} \nonumber\\
		\, &= \, - \, \frac{\pi^2 (q-1)B_1b^{q-2}}{3q\beta^2} \int dt_3 dt_4 \left[ b^q \left( \frac{{\rm sgn}(t_{13}){\rm sgn}(t_{24})}{|t_{13}|^{-\frac{2}{q}}|t_{24}|^{2-\frac{2}{q}}}
		+ \frac{{\rm sgn}(t_{13}){\rm sgn}(t_{24})}{|t_{13}|^{2-\frac{2}{q}}|t_{24}|^{-\frac{2}{q}}} \right) 
		+ (q-2) \frac{\delta(t_{13})\delta(t_{24})}{|t_{12}|^{-\frac{4}{q}}} \right] \frac{{\rm sgn}(t_{34})}{|t_{34}|^{\frac{2}{q}+1}}\, .
	\end{align}
Again one can evaluate the integrals and find $c_{1,2}=-(q-1)\pi^2/3q$.
Finally we consider $\mathcal{O}(\beta^{-3})$ order. The equation of this order reads
	\begin{align}
		B_1 c_{1,3} \beta^{-3} \int dt_3 dt_4 \, \mathcal{K}(t_1, t_2; t_3, t_4) \frac{{\rm sgn}(t_{34})}{|t_{34}|^{\frac{2}{q}-2}} \, = \, 0 \, .
	\end{align}
The LHS integral identically vanishes. Hence, we cannot determine the coefficient $c_{1,3}$ from this equation.
Nevertheless, this iterative method precisely recovers the expansion of (\ref{M/S solution}) up to the third order:
	\begin{align}
		\delta G(t_1, t_2) \, = \, - \, B_1 \ \frac{{\rm sgn}(t_{12})}{|t_{12}|^{\frac{2}{q}+1}} \left[ 1 - \frac{(q-1)\pi^2}{3q} \left| \frac{t_{12}}{\beta} \right|^2
		+ \frac{2\pi^2}{3} \left| \frac{t_{12}}{\beta} \right|^3 - \frac{(2q-1)(q+5)\pi^4}{90q^2} \left| \frac{t_{12}}{\beta} \right|^4 + \cdots \right] \, .
	\label{MS-expansion}
	\end{align}
where we used the relation (\ref{alpha_G}).
Using this $\Psi_{1, \beta}$ expansion as source together with $\Psi_{0, \beta}$, we can also apply this method to determine low temperature expansion of $\Psi_{2, \beta}$.

\subsection{Tree-Level Free Energy}
\label{sec:tree-level free energy}
Now we evaluate the tree-level free energy through the regularized breaking term.
The order $(\beta J)^0$ contribution to the tree-level free energy, which comes from $S_{\rm c}[\Psi_{0, \beta}]$, was already evaluated in \cite{Kitaev:2015,Maldacena:2016hyu,Jensen:2016pah}.
Therefore in this section, we will evaluate higher order contributions of the $1/\beta J$ expansion to the tree-level free energy.

\subsubsection{$(\beta J)^{-1}$ Contribution}
\label{sec:beta J^-1 contribution}
The action of the collective time coordinate was evaluated in Appendix \ref{app:s-regularization and schwarzian action} from the regularized breaking term,
which leads to the Schwarzian action given in Eq.(\ref{S[f]}).
Now, we use the classical solution: $f(t)=\frac{\beta}{\pi}\tan(\frac{\pi t}{\beta})$.
Then, the integral can be evaluated to give $2\pi^2/\beta$.
Therefore, the $S[f]$ contribution to the tree-level free energy is 
	\begin{equation}
		\beta F_0 \, = \, \frac{NB_1 \gamma \pi^2}{\beta J} \, .
	\label{first order F}
	\end{equation}
This contribution can actually be evaluated directly from the regularized breaking term by
	\begin{equation}
		\beta F_0 \, = \, - \, \frac{N}{2} \lim_{s \to \frac{1}{2}} \int dt_1 dt_2 \, \Psi_{0,\beta}(t_1, t_2) \, Q_s(t_1, t_2) \, ,
	\end{equation}
where the finite temperature critical solution $\Psi_{0, \beta}$ and the regularized source $Q_s$ are given in Eq.(\ref{Psi_0-beta}) and (\ref{Q_s}), respectively.
Since the regularized source $Q_s$ has a factor $(s-1/2)$, in order to obtain non-vanishing contribution after the limit,
we only need to extract a single pole $(s-1/2)^{-1}$ term from the integral.
For this purpose, we expand the finite temperature critical solution $\Psi_{0, \beta}$ by power series of $|t_{12}|$ up to $|t|^{2-\frac{2}{q}}$ order,
which is responsible for a single pole term.
This leads to 
	\begin{equation}
		\beta F_0 \, = \, - \, \frac{NB_1 \pi^2}{6 q\beta J} \, (q-1) b^{q-1} \left[ \gamma(s, q) \int \frac{dt}{|t|^{2s}} \right]_{s \to \frac{1}{2}} \, .
	\end{equation}
Hence, using the expansion of $\gamma(s,q)$ in Eq.(\ref{gamma(s,q)-expansion}) and taking the limit $s \to 1/2$, we obtain the final result.
This result agrees with the result found in Eq.(\ref{first order F}) from the Schwarzian action.

\subsubsection{$(\beta J)^{-2}$ Contribution}
\label{sec:beta J^-2 contribution}
Now we consider the next $(\beta J)^{-2}$ order contribution.
The contribution from the breaking term to such order is given by
	\begin{equation}
		\beta F_1 \, = \, - \, \frac{N}{2} \lim_{s \to \frac{1}{2}} \int dt_1 dt_2 \, \Psi_{1,\beta}(t_1, t_2) \, Q_s(t_1, t_2) \, .
	\label{second order F}
	\end{equation}
Again to compute this free energy, we only need to extract the $|t|^{2-\frac{2}{q}}$ order term from $\Psi_{1,\beta}$.
From the expansion in Eq.(\ref{MS-expansion}), one can read off the $|t|^{2-\frac{2}{q}}$ order term as
	\begin{equation}
		\Psi_{1,\beta}(t_1, t_2) \, = \, - \, \frac{2B_1 \pi^2}{3\beta^3 J^{1+\frac{2}{q}}} \, \frac{{\rm sgn}(t_{12})}{|t_{12}|^{\frac{2}{q}-2}} \, + \, \cdots \, .
	\end{equation}
Following the same process as in the previous subsection, one can evaluate the contribution from the breaking term to this order free energy. 
However, this is not the all contributions to this $(\beta J)^{-2}$ order free energy.
The critical action $S_{\rm c}$ part also gives a contribution to this $(\beta J)^{-2}$ order, which is half of the breaking term contribution with opposite sign. 
Therefore, combining these two contributions, the final answer for the $(\beta J)^{-2}$ order free ernrgy is given by
	\begin{equation}
		\beta F_1 \, = \, - \, \pi^2 q \ \frac{NB_1^2\gamma}{b(\beta J)^2} \, .
	\end{equation}

\subsubsection{$(\beta J)^{-n}$ Contribution}
\label{sec:beta J^-n contribution}
In this subsection, we discuss the general $(\beta J)^{-n}$ order contribution of the tree-level free energy.
For this purpose, let us first look at the collective action (\ref{S_col[Psi, f]}).
After rescaling the bi-local field by $\Psi \to J^{-2/q} \Psi$, one sees the explicit $J$-dependence appearing only in the breaking term.
Hence, from the $J$-derivative trick, the tree-level free energy is solely determined by the breaking term by
	\begin{equation}
		J \frac{\partial}{\partial J} \big( \beta F_n \big) \, = \, \frac{N}{2J} \lim_{s \to \frac{1}{2}} \int dt_1 dt_2 \, \Psi_{n,\beta}(t_1, t_2) Q_s(t_1, t_2) \, .
	\label{derivative of F}
	\end{equation}
We know that any order of $1/J$ correction for the zero temperature classical solution is given by Eq.(\ref{Psi_n}).
Even though we don't know exact finite-temperature version of these corrections, we nevertheless expect the finite-temperature solution can be expanded in low temperature region as
	\begin{equation}
		\Psi_{n, \beta}(t_1, t_2) \, = \, \frac{B_n}{J^n} \frac{{\rm sgn}(t_{12})}{|t_{12}|^{\frac{2}{q}+n}} \, \left[ 1 + \cdots + a_n \left| \frac{t_{12}}{\beta} \right|^{n+2} + \cdots \right] \, ,
	\end{equation}
where $a_n$ is a $q$-dependent constant, but independent of $J$, $\beta$ or $t$.
As we saw in the previous sections, the $|t|^{2-\frac{2}{q}}$ order term is only needed to extract the $(s-1/2)^{-1}$ poles.
Hence, substituting this order term into Eq.(\ref{derivative of F}), one can perform the integrals and the limit together with Eq.(\ref{Q_s}).
This result is given by
	\begin{equation}
		J \frac{\partial}{\partial J} \big( \beta F_n \big) \, = \, - \, \frac{3qa_n}{b} \, \frac{NB_1B_n \gamma}{(\beta J)^{n+1}} \, .
	\end{equation}
After the integration of $J$, the free energy is given by
	\begin{equation}
		\beta F_n \, = \, \frac{3qa_n}{(n+1)b} \, \frac{NB_1B_n \gamma}{(\beta J)^{n+1}} \, .
	\end{equation}
We can check the consistency of this formula with previous results.
For $n=0$, we have $B_0=b$ and $a_0=\pi^2/(3q)$.
Then the formula gives the result we found in Section \ref{sec:beta J^-1 contribution}.
For $n=1$, we have $a_1=-2\pi^2/3$, and then the formula again leads to the result found in Section \ref{sec:beta J^-2 contribution}.
For general order we only need to determine $a_n$ to evaluate the free energy.
We note that in principle the coefficient of the zero temperature solution $B_n$ can be determined from the recursion relation (\ref{recursion relation}).

In summary, we have obtained the following $1/J$ corrections to the tree-level free energy
	\begin{equation}
		\frac{\beta F}{N} \, = \, \frac{B_1 \gamma \pi^2}{\beta J} \, - \, \frac{B_1^2 \gamma \pi^2 q}{b(\beta J)^2}
		\, - \, \frac{3qa_2 B_1^3 \gamma}{b^2(\beta J)^3} \left( \frac{q+2}{8q} \right) \left[ (q-2) + (3q-2) \tan^2\left( \tfrac{\pi}{q}\right) \right] \, + \, \cdots \, .
	\end{equation}
For $q=4$, we can compute the coefficients as
	\begin{equation}
		\frac{\beta F}{N} \, = \, - \, 0.197 \, (\beta J)^{-1} \, + \, 0.208 \, (\beta J)^{-2} \, + \, 0.038 \times a_2 \, (\beta J)^{-3} \, + \, \cdots \, ,
	\end{equation}
with $a_2$ to be determined.
These results agree with the recent numerical results of \cite{Garcia-Garcia:2016mno,Cotler:2016fpe}.

\section{Conclusion}
\label{sec:conclusion}
In the present paper we have completed the formulation given in (I) in defining a reparametrization invariant collective theory at the IR point and away from it.
A regularized action representing an interacting theory between a Schwarzian coordinate and bi-local matter is specified.
It generates perturbative calculations in the SYK model around the conformal IR point which are systematic in the inverse of the strong coupling $J$.
We gave the evaluation of the tree level free energy in this expansion.
Even though, the present calculations are done at tree level in $1/N$, the formalism given allows for loop level calculations with no difficulty:
by projection of the zero mode the perturbation expansion is well defined,
while the Jacobian(s) of the changes of variables provide exact counter terms which are expected to cancel infinities appearing in loop diagrams.

These higher order calculations and further detailed study of the model will be of definite usefulness regarding 
the question of the exact AdS$_2$ Gravity dual representing this theory.
A class of dilation Gravities related to the models developed by Almheiri and Polchinski \cite{Almheiri:2014cka} shows features contained in SYK model
\cite{Jensen:2016pah,Maldacena:2016upp,Engelsoy:2016xyb,Grumiller:2016dbn}.
The representation that we have given with exact action featuring interaction between the dynamical (time) coordinate and bi-local matter
is the system that one might hope to recover from the corresponding AdS$_2$ theory.

\acknowledgments
We acknowledge useful conversation with Robert de Melo Koch and Sumit Das on the topics of this paper.
This work is supported by the Department of Energy under contract DE-SC0010010. 
Most importantly, we are grateful to Stephen Shenker for pointing out problems with our first evaluation of the coefficient governing the nonlinear Schwarzian action.
The improved, regularized evaluation that followed is a consequence of this discussion.

\appendix

\section{$\epsilon$-Expansion}
\label{app:epsilon-expansion}
In this appendix, we will exemplify how to obtain the non-linear Schwarzian action (\ref{S[f]}) associated with the naive form of the breaking term in Eq.(\ref{S_col}).
This is done by using $\varepsilon$-expansion with $q=2/(1-\varepsilon)$ and treating $\varepsilon$ as a small parameter.
We note that for any $q$ in the range of $2\le q \le \infty$, the value of $\varepsilon$ is $0\le \varepsilon \le 1$.
Therefore, the convergence of this $\epsilon$-expansion is guaranteed.
Even though we use the $\varepsilon$-expansion, we can nevertheless calculate all order contributions of $\varepsilon$ as we will see below.
We first rewrite the critical solution in the following way:
	\begin{align}
		\Psi_{0, f}(t_1, t_2) \, &= \, - \, \frac{1}{\pi J} \left( \frac{\sqrt{|f'(t_1)f'(t_2)|}}{|f(t_1) -f(t_2)|} \right) \nonumber\\
		&\hspace{30pt} \times \left[ \, 1 \, - \, \varepsilon \, \log\left( \frac{\sqrt{|f'(t_1)f'(t_2)|}}{|f(t_1) -f(t_2)|} \right)
		\, + \, \frac{\varepsilon^2}{2} \, \left( \log\frac{\sqrt{|f'(t_1)f'(t_2)|}}{|f(t_1) -f(t_2)|} \right)^2 \, + \, \cdots \right] \, ,
	\end{align}
where the first term is the contribution from $q=2$ case, which leads to the result Eq.(\ref{S[f]}) with $\alpha=1$.
To evaluate higher order $\varepsilon$ contributions, we use the following expansions of the logarithm in the $t_1 \to t_2$ limit: 
	\begin{equation}
		\log\left( \frac{\sqrt{|f'(t_1)f'(t_2)|}}{|f(t_1) -f(t_2)|} \right)
		\, = \, - \, \log|t_1-t_2| \, - \, \frac{1}{8} \frac{|f''(t_2)|^2}{|f'(t_2)|^2} |t_1-t_2|^2 \, + \, \frac{1}{12} \frac{|f'''(t_2)|}{|f'(t_2)|} |t_1-t_2|^2 \, + \, \cdots \, .
	\label{log expansion}
	\end{equation}
The first log term gives an $f$-independent divergent term which we will eliminate in the following.
One also expands the factor representing $q=2$ reparametrized critical solution and then one finds $\mathcal{O}(\varepsilon) = 0$.
For order $\mathcal{O}(\varepsilon^2)$ contribution, from Eq.(\ref{log expansion}), one can find
	\begin{align}
		\mathcal{O}(\varepsilon^2) \, &= \, - \, \frac{N\varepsilon^2}{4\pi J} \int dt_1 \, \partial_1
		\left[ \left( \frac{1}{4} \frac{|f''(t_2)|^2}{|f'(t_2)|^2} \, - \, \frac{1}{6} \frac{|f'''(t_2)|}{|f'(t_2)|} \right) |t_1-t_2| \, \log|t_1-t_2| \right]_{t_2 \to t_1} \nonumber\\
		&= \, \frac{N\varepsilon^2}{24\pi J} \int dt_1 \, \left[ \, \frac{f'''(t_1)}{f'(t_1)} \, - \, \frac{3}{2} \left( \frac{f''(t_1)}{f'(t_1)} \right)^2 \, \right] \, ,
	\end{align}
where we again eliminated the divergence term and used integration by parts.
Hence, the total contribution up to $\mathcal{O}(\varepsilon^2)$ for $q=2/(1-\varepsilon)$ action is given by 
	\begin{equation}
		S[f] \, = \, - \, \frac{N\alpha}{24\pi J} \int dt \, \left[ \, \frac{f'''(t)}{f'(t)} \, - \, \frac{3}{2} \, \left( \frac{f''(t)}{f'(t)} \right)^2 \, \right] \, ,
	\end{equation}
where 
	\begin{equation}
		\alpha(\varepsilon) \, = \, 1 \, - \, \varepsilon^2 \, + \, \mathcal{O}(\varepsilon^3) \, .
	\label{alpha}
	\end{equation}
In fact, there is no higher order contributions from $\mathcal{O}(\varepsilon^3)$. This can be seen from an expansion
	\begin{align}
		&\quad \ \left( \log \frac{\sqrt{|f'(t_1)f'(t_2)|}}{|f(t_1) -f(t_2)|} \right)^n \nonumber\\
		\, &= \, \Big( - \log|t_1-t_2| \Big)^n
		\, - \, n \left( \frac{1}{8} \frac{|f''(t_2)|^2}{|f'(t_2)|^2} \, - \, \frac{1}{12} \frac{|f'''(t_2)|}{|f'(t_2)|} \right) |t_1-t_2|^2 \, \Big( -\log|t_1-t_2| \Big)^{n-1} \, + \, \cdots \, .
	\end{align}
This expansion together with the expansion of $q=2$ reparametrized critical solution does not give any non-zero finite contribution to the action after the limit when $n\ge3$.
Namely, the $(\log|t_1-t_2|)^n$ factor gives a strong divergence when $n$ is large.
However, if one wants to lower the power of this logarithm, then one gets a higher power of $|t_1-t_2|^n$, which strongly vanishes after setting $t_2=t_1$.
This naive form of $\alpha$ will turn out to be renormalized with our regularization of the breaking term.
We evaluate this renormalized coefficient in the next appendix.

\section{$s$-Regularization and Schwarzian Action}
\label{app:s-regularization and schwarzian action}
In this appendix, we will directly evaluate the collective coordinate action with the regularized breaking term:
	\begin{equation}
		S[f] \, = \, \frac{N}{2} \int \, \big[ \Psi_{0, f} \big]_s
		\, = \, - \, \frac{N}{2} \lim_{s \to \frac{1}{2}} \int dt_1 dt_2 \, \Psi_{0, f}(t_1, t_2) \, Q_s(t_1, t_2) \, ,
	\label{S[f]-app:reg}
	\end{equation}
and confirm that the result is given by 
	\begin{equation}
		S[f] \, = \, - \, \frac{N\alpha}{24\pi J} \int dt \, \left[ \, \frac{f'''(t)}{f'(t)} \, - \, \frac{3}{2} \, \left( \frac{f''(t)}{f'(t)} \right)^2 \, \right] \, ,
	\end{equation}
with a coefficient 
	\begin{equation}
		\alpha \, = \, - 12 \pi B_1 \gamma \, .
	\end{equation}

For this purpose, we expand the reparametrized critical solution with $f(t)=t+\varepsilon(t)$ as
	\begin{align}
		\Psi_{0,f}(t_1, t_2) \, &= \, \Psi_0(t_{12}) \, + \, \int dt_a \, \varepsilon(t_a) \, u_{0, t_a}(t_1, t_2) 
		 \, + \, \frac{1}{2} \int dt_a dt_b \, \varepsilon(t_a) \varepsilon(t_b) \, u_{(1), t_a, t_b}(t_1, t_2) \nonumber\\
		&\qquad + \, \frac{1}{6} \int dt_a dt_b dt_c \, \varepsilon(t_a) \varepsilon(t_b) \varepsilon(t_c) \, u_{(2), t_a, t_b, t_c}(t_1, t_2) \, + \, \cdots \, ,
	\label{Psi_{0, f} expansion}
	\end{align}
where we defined
	\begin{align}
		u_{0, t_a}(t_1, t_2) \, &\equiv \, \frac{\partial \Psi_{0, f}(t_{12})}{\partial f(t_a)} \Bigg|_{f(t)=t} \, , \nonumber\\
		u_{(2), t_a, t_b}(t_1, t_2) \, &\equiv \, \frac{\partial^2 \Psi_{0, f}(t_{12})}{\partial f(t_a) \partial f(t_b)} \Bigg|_{f(t)=t} \, , \nonumber\\
		u_{(3), t_a, t_b, t_c}(t_1, t_2) \, &\equiv \, \frac{\partial^3 \Psi_{0, f}(t_{12})}{\partial f(t_a) \partial f(t_b) \partial f(t_c)} \Bigg|_{f(t)=t} \, , \nonumber\\
		u_{(n), \{t_a, \cdots, t_n\}}(t_1, t_2) \, &\equiv \, \frac{\partial^n \Psi_{0, f}(t_{12})}{\partial f(t_a) \cdots \partial f(t_n)} \Bigg|_{f(t)=t} \, .
	\end{align}
Let us first consider the quadratic and cubic order contributions.
Taking derivatives and expressing in the momentum space, the quadratic and cubic coefficients are given by
	\begin{align}
		u_{(2), \omega_a, \omega_b}(t_1, t_2)
		&= \, \frac{2}{q} \ e^{i (\omega_a+\omega_b)t_+} \left[ |\omega_a \omega_b| \cos((\omega_a+\omega_b)t_-) \, - \, \frac{1}{|t_-|^2} \sin|\omega_a t_-| \sin|\omega_b t_-| \right]
		 \, \Psi_0(t_{12}) \nonumber\\
		&\, - \, \frac{2}{\pi q^2} \ e^{i (\omega_a+\omega_b)t_+} \, |\omega_a \omega_b|^{\frac{1}{2}} |t_-| \, J_{\frac{3}{2}}(|\omega_a t_-|) \, J_{\frac{3}{2}}(|\omega_b t_-|) \, \Psi_0(t_{12}) \, , 
	\end{align}
and 
	\begin{align}
		&\quad u_{(3), \omega_a, \omega_b, \omega_c}(t_1, t_2) \nonumber\\
		&= \, - \, \frac{4i}{q} \ e^{i (\omega_a+\omega_b+\omega_c)t_+} \left[ |\omega_a \omega_b \omega_c| \cos((\omega_a+\omega_b+\omega_c)t_-)
		\, - \, \frac{1}{|t_-|^3} \sin|\omega_a t_-| \sin|\omega_b t_-| \sin|\omega_c t_-| \right] \, \Psi_0(t_{12}) \nonumber\\
		&\, - \, \frac{4i}{q^2} \ \sqrt{\frac{\pi |\omega_c^3 t_-|}{2}} e^{i (\omega_a+\omega_b+\omega_c)t_+}
		\left[ |\omega_a \omega_b| \cos((\omega_a+\omega_b)t_-) \, - \, \frac{1}{|t_-|^2} \sin|\omega_a t_-| \sin|\omega_b t_-| \right] \, J_{\frac{3}{2}}(|\omega_c t_-|) \, \Psi_0(t_{12}) \nonumber\\
		&\, + \, \frac{8i}{q^3} \, \left| \frac{\pi t_-}{2} \right|^{\frac{3}{2}} e^{i (\omega_a+\omega_b+\omega_c)t_+} \,
		|\omega_a \omega_b\omega_c|^{\frac{3}{2}} \, J_{\frac{3}{2}}(|\omega_a t_-|) \, J_{\frac{3}{2}}(|\omega_b t_-|) \, J_{\frac{3}{2}}(|\omega_c t_-|) \, \Psi_0(t_{12}) \, .
	\end{align}
In fact, there are two more terms in the second line of RHS in $u_{(3)}$ obtained by permutations of $(\omega_a, \omega_b, \omega_c)$, but we omitted these terms in the above expression.
Substituting these expressions into the action (\ref{S[f]-app:reg}) and performing the $t_1$, $t_2$ integrals,
one finds single poles $(s-1/2)^{-1}$ coming from the double sine term in $u_{(2)}$ and from the triple sine term in $u_{(3)}$.
Namely for the quadratic contribution, we have
	\begin{align}
		&- \, \int dt_1 dt_2 dt_3 dt_4 \, u_{(2), \omega_a, \omega_b}(t_1, t_2) \, \mathcal{K}(t_1, t_2; t_3, t_4) \Psi_1(t_3, t_4) \nonumber\\
		&= \, 2^{2-2s} \pi \left( \frac{q-1}{q} \right) B_1 b^{q-1} \gamma(s, q) \, \delta(\omega_a+\omega_b) \int_0^{\infty} \frac{dt_-}{|t_-|^{4+2s}} \, \sin^2|\omega_a t_-| \, ,
	\end{align}
with
	\begin{equation}
		\int_0^{\infty} dx \ \frac{\sin^2x}{x^{4+2s}} \, = \, \frac{1}{6} \, \frac{1}{s-\tfrac{1}{2}} \, + \, \mathcal{O}((s-\tfrac{1}{2})^0) \, .
	\end{equation}
Also for the cubic contribution
	\begin{align}
		&- \, \int dt_1 dt_2 dt_3 dt_4 \, u_{(3), \omega_a, \omega_b, \omega_c}(t_1, t_2) \, \mathcal{K}(t_1, t_2; t_3, t_4) \Psi_1(t_3, t_4) \nonumber\\
		&= \, - \, 2^{3-2s} \pi i \left( \frac{q-1}{q} \right) B_1 b^{q-1} \gamma(s, q) \, \delta(\omega_a+\omega_b+\omega_c)
		\int_0^{\infty} \frac{dt_-}{|t_-|^{5+2s}} \, \sin|\omega_a t_-| \, \sin|\omega_b t_-| \, \sin|\omega_c t_-| \, ,
	\end{align}
with
	\begin{align}
		&\quad \int_0^{\infty} \frac{dt_-}{|t_-|^{5+2s}} \, \sin|\omega_a t_-| \, \sin|\omega_b t_-| \, \sin|\omega_c t_-| \nonumber\\
		\, &= \, |\omega_a \, \omega_b \, \omega_c| \, \big( |\omega_a|^2+|\omega_b|^2+|\omega_c|^2 \big) \, \frac{1}{12(s-\tfrac{1}{2})} \, + \, \mathcal{O}((s-\tfrac{1}{2})^0) \, .
	\end{align}
There are no other terms giving such $(s-1/2)^{-1}$ pole.
Such single pole factor $(s-1/2)^{-1}$ cancels with the $(s-1/2)$ factor in the regularized source $Q_s$ (\ref{Q_s}), and lead to 
	\begin{equation}
		- \, \int dt_1 dt_2 dt_3 dt_4 \, u_{(2), t_a, t_b}(t_1, t_2) \, \mathcal{K}(t_1, t_2; t_3, t_4) \Psi_1(t_3, t_4)
		\, = \, B_1 \, \gamma \, \partial_{t_a}^2 \partial_{t_b}^2 \delta(t_{ab}) \, ,
	\end{equation}
and 
	\begin{align}
		&- \, \int dt_1 dt_2 dt_3 dt_4 \, u_{(3), t_a, t_b, t_c}(t_1, t_2) \, \mathcal{K}(t_1, t_2; t_3, t_4) \Psi_1(t_3, t_4) \nonumber\\
		\, &= \, B_1 \, \gamma \, \partial_{t_a} \partial_{t_b} \partial_{t_c} \big( \partial_{t_a}^2 + \partial_{t_b}^2 + \partial_{t_c}^2 \big) \, \delta(t_{ac}) \delta(t_{bc}) \, .
	\end{align}

With the experience of quadratic and cubic order computations, now we would like to evaluate all order contributions.
As we saw above the poles associated to the limit $s \to 1/2$ only come from the double and triple sine terms.
Therefore, we expect this structure is also true for any higher order contributions.
Taking derivatives of the reparameterized  critical solution, we find such term in $n$-th order is given by
	\begin{equation}
		u_{(n), \{\omega_a, \cdots, \omega_n\}}(t_1, t_2) \, = \, \frac{2}{q} (-i)^n (n-1)! \, e^{i(\omega_a + \cdots + \omega_n) t_+} \, \left( \prod_{i=a}^n \sin|\omega_i t_-| \right)
		\frac{\Psi_0(t_{12})}{|t_-|^n} \, + \, \cdots \, ,
	\end{equation}
where the ellipsis denotes non-singular terms in the limit $s \to 1/2$.
Now, using the result (\ref{Q_s}), one obtains the contribution from the $n$-th order as
	\begin{align}
		&- \, \int dt_1 dt_2 dt_3 dt_4 \, u_{(n), \{\omega_a, \cdots, \omega_n\}}(t_1, t_2) \, \mathcal{K}(t_1, t_2; t_3, t_4) \Psi_1(t_3, t_4) \nonumber\\
		&= \, - \, 12 \pi (-i)^n (n-1)! \, \gamma B_1 \, (s - \tfrac{1}{2}) \delta(\omega_a + \cdots + \omega_n) \int_0^{\infty} \frac{dt_-}{|t_-|^{2+2s+n}}
		\left( \prod_{i=a}^n \sin|\omega_i t_-| \right)
		\, + \, \mathcal{O}(s-\tfrac{1}{2}) \, .
	\end{align}
The $t_-$-integral is given by
	\begin{equation}
		\quad \int_0^{\infty} \frac{dt_-}{|t_-|^{2+2s+n}} \, \prod_{i=1}^n \, \sin|\omega_i t_-|
		\, = \, \left( \prod_{i=1}^n |\omega_i| \right) \, \left( \sum_{i=1}^n |\omega_i|^2 \right) \, \frac{1}{12(s-\tfrac{1}{2})} \, + \, \mathcal{O}((s-\tfrac{1}{2})^0) \, .
	\end{equation}
Now, using this result and Fourier transforming back to $\{t_a, \cdots, t_n\}$ from $\{\omega_a, \cdots, \omega_n\}$, we get
	\begin{align}
		&- \, \int dt_1 dt_2 dt_3 dt_4 \, u_{(n), \{t_a, \cdots, t_n\}}(t_1, t_2) \, \mathcal{K}(t_1, t_2; t_3, t_4) \Psi_1(t_3, t_4) \nonumber\\
		&= \, \frac{(n-1)!}{2} \, B_1 \gamma \, \left( \prod_{i=a}^n \partial_{t_i} \right) \left( \sum_{i=a}^n \partial_{t_i}^2 \right) \delta(t_{an}) \cdots \delta(t_{n-1, n}) \, , 
	\end{align}
where we have already taken $s\to 1/2$ limit.

Finally, together with the expansion (\ref{Psi_{0, f} expansion}), one can see that the $n$-th order contribution to the collective coordinate action (\ref{S[f]}) is given by
	\begin{align}
		S[f] \, &= \, \frac{NB_1 \gamma}{4nJ} \, \int dt_1 \cdots dt_n \, \varepsilon(t_1) \cdots \varepsilon(t_n) 
		\left( \prod_{i=1}^n \partial_{t_i} \right) \left( \sum_{i=1}^n \partial_{t_i}^2 \right) \delta(t_{1n}) \cdots \delta(t_{n-1, n}) \nonumber\\
		&= \, \frac{N B_1 \gamma}{2J} \int dt \, \frac{(-1)^n}{2} \varepsilon'''(t) \Big( \varepsilon'(t) \Big)^{n-1} \, .
	\label{S[f]-2}
	\end{align}
This result can be summed over for all order to get
	\begin{equation}
		S[f] \, = \, \frac{NB_1 \gamma}{2J} \int dt \, {\rm Sch}(f; t) \, ,
	\end{equation}
where 
	\begin{equation}
		{\rm Sch}(f; t) \, = \, \frac{f'''(t)}{f'(t)} \, - \, \frac{3}{2} \left( \frac{f''(t)}{f'(t)} \right)^2 \, .
	\end{equation}
To see this correspondence, one first rewrites the Schwarzian derivative by integration by parts as
	\begin{equation}
		\int dt \, {\rm Sch}(f; t) \, = \, - \frac{1}{2} \frac{f'''(t)}{f'(t)} \, .
	\end{equation}
Then, we use $f(t)=t+\varepsilon(t)$ and expand the Schwarzian derivative by powers of $\varepsilon$ as
	\begin{equation}
		\int dt \, {\rm Sch}(f; t) \, = \, \int dt \, \sum_{n=1}^{\infty} \frac{(-1)^n}{2} \, \varepsilon'''(t) \Big( \varepsilon'(t) \Big)^{n-1} \, .
	\end{equation}
This expansion completely agrees with the result found in Eq.(\ref{S[f]-2}).

Finally as a reference, we give a relation of our coefficients to the coefficients $\alpha_S$ and $\alpha_G$ defined in \cite{Maldacena:2016hyu}:
	\begin{equation}
		- \, \frac{\alpha}{12\pi} \, = \, B_1 \gamma \, = \, - \, 2 \alpha_S \left( \frac{J}{\mathcal{J}} \right) \, = \, b \, \gamma \, \alpha_G \left( \frac{J}{\mathcal{J}} \right) \, ,
	\end{equation}
where $\mathcal{J}=\frac{\sqrt{q}}{2^{\frac{q-1}{2}}}J$.


\end{document}